\documentclass[twocolumn,showpacs,amsmath,amssymb]{revtex4}
\usepackage{graphicx}
\newcommand{\ba}{\begin{eqnarray}}
\newcommand{\ea}{\end{eqnarray}}

\newcommand{\beq}{\begin{equation}}
\newcommand{\eeq}{\end{equation}}
\newcommand{\beqs}{\begin{eqnarray}}
\newcommand{\eeqs}{\end{eqnarray}}

\newfont{\prg}{cmsy10}

\begin{document}

\title{ How precisely will the total cross section be measured at the LHC?  }

\author{J.-R. Cudell$^1$\thanks{JR.Cudell@ulg.ac.be} and
O.~V. Selyugin$^{2}$ \thanks{selugin@thsun1.jinr.ru}}

\affiliation{  $^1$ \it IFPA, AGO Department B5a, Universit\'e de Li\`ege, 4000
Li\`ege, Belgium;  \\ $^2$ \it BLTP,
Joint Institute for Nuclear Research,
141980 Dubna, Moscow region, Russia}

\begin{abstract}
It is very likely that hadronic scattering will enter a new regime at the
LHC, as the black-disk limit is reached. This will lead to
a severe change in the $t$ dependence of the real part and of the slope of the 
elastic scattering 
amplitude, and in turn this may bias the measurement of the total cross section.
We examine this issue, and suggest new strategies to test the reliability
of the total cross section measurements.
\end{abstract}
\maketitle

Many models predict that soft interactions will enter a new regime at the LHC:
 given the huge energy, unitarisation may play a crucial role as the central 
part of the protons becomes black. Indeed, all
simple-power fits to lower energy data will violate unitarity in some partial-wave 
amplitudes before the LHC energy. 
This means that  something will happen that will restore it $-$ call it saturation, 
unitarisation or emergence of cuts $-$ and this will modify the expectations one has 
from Regge models based on simple poles. 
While conventional models predict a total cross section from 90 to 125~mb 
\cite{COMPETE}, the presence of
of a hard pomeron gives around 150 mb \cite{physlett,SelyuginBL07,DoLa} and 
U-matrix  unitarisation can give 230 mb \cite{Troshin}. This clearly shows that the
uncertainties due to the underlying models, and especially due to the unitarisation 
scheme, are very large. One may hope to select the true models
by a measurement of the total cross section.

The LHC will be well equipped
to study in depth the diffractive processes, as it will have a superb rapidity coverage, 
and two experiments $-$ TOTEM \cite{TOTEM} and ATLAS 
\cite{ATLAS} $-$ plan to measure the total cross section.
They intend to reach an accuracy on $\sigma_{tot}$ of the order of
$1 \% $, which would indeed give a very stringent test of theory.
A few assumptions underlie this estimate of the accuracy. First of all, the 
ratio $\rho$ of the real part to the imaginary part of the elastic scattering amplitude,
is assumed to be small, and to vary little with $t$: $\rho(s,t)\approx 0.14$ \cite{COMPETE}. Secondly, the elastic cross section is assumed to fall exponentially
with $t$: $dN/dt\sim \exp(Bt)$. 

We want to show here that, if the elastic $p p$ amplitude reaches a new 
regime at the LHC, it will invalidate the above assumptions and  the 
measurement of the total cross section will be biased and much more uncertain
than foreseen. 
Indeed, most unitarisation schemes lead to
novel properties of the elastic amplitude. The problems that we talk about here 
concern a large class of models, in which the elastic amplitude contains a 
fast-rising component that needs to be unitarised. We have checked that they are present
in models saturating the profile function, in models using analytic
unitarisation schemes \cite{physlett} or in the Dubna Dynamical Model \cite{DDM}.

To illustrate our point, we shall consider a simple unitarised two-component model, which
includes a soft pomeron and a hard pomeron, and which we shall call the eikonalised 
two-pomeron model (ETPM). This model 
is based on a fit to soft data which includes a hard pomeron
component \cite{CMSL} of intercept 1.4 that accounts for the growth of the
gluon density at small $x$ in deeply inelastic processes \cite{LandshoffHP3}.  
Although the coefficient of the hard pomeron term is small in soft data, it grows like 
$s^{0.4}$ so
that the amplitude will reach the black-disk limit 
at small impact parameter $b$ before the LHC energy
 \cite{SelyuginBDLCJ04,physlett}. The amplitude must then be unitarised and, 
in this simple model, we consider a 1-channel eikonal in the impact-parameter 
representation \cite{physlett}.
The net effect of this unitarisation is to make 
$B(s,t)$ increase with $|t|$ at small $|t|$ for LHC energies, as shown in Fig. 1.
We also show in that figure that  the $t$-dependence of $\rho(s,t)$ changes
drastically.
\begin{figure}[htp]
\vglue -1cm
\includegraphics[width=0.4\textwidth] {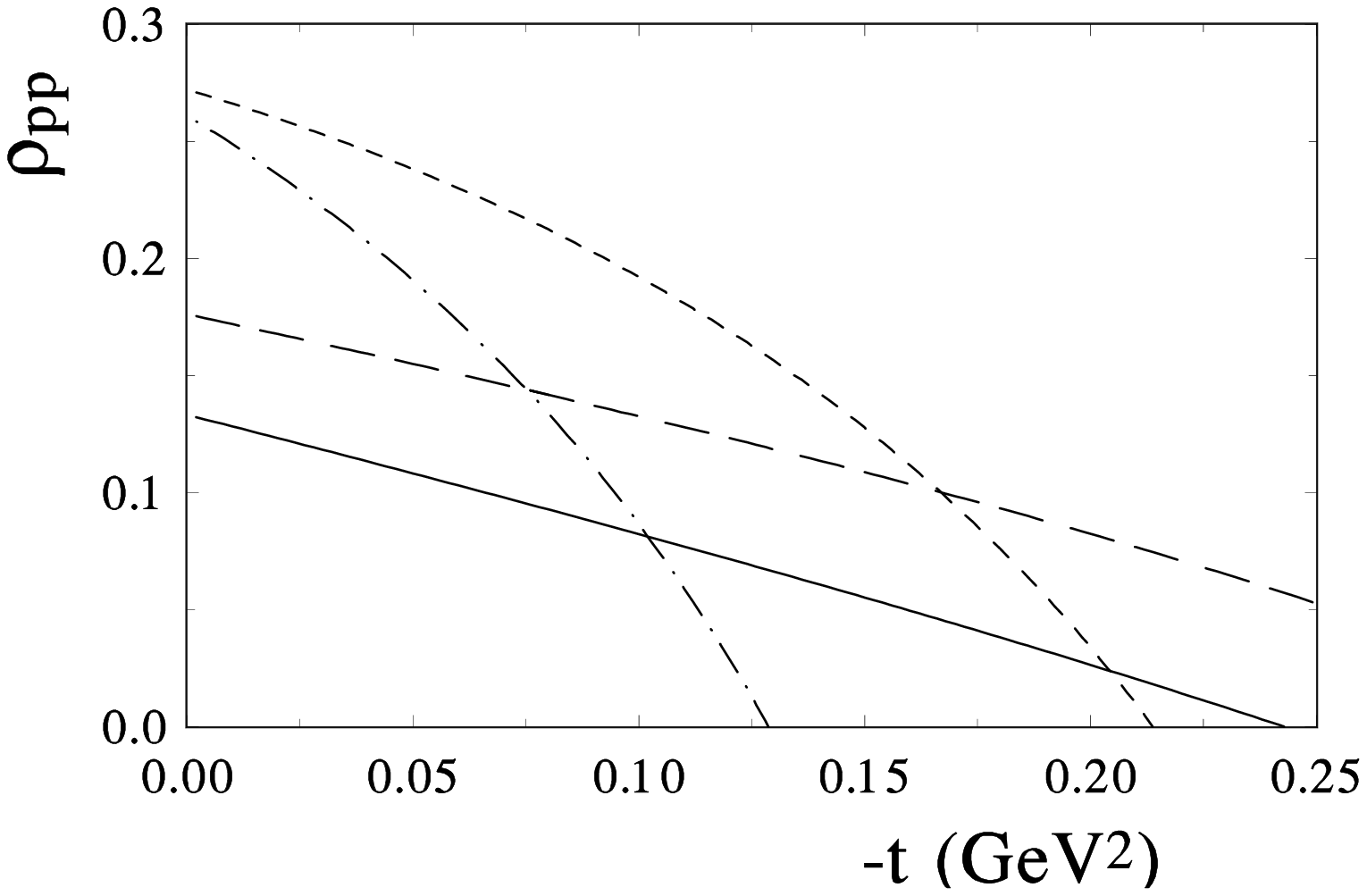}\\
\vglue -1cm\includegraphics[width=0.4\textwidth] {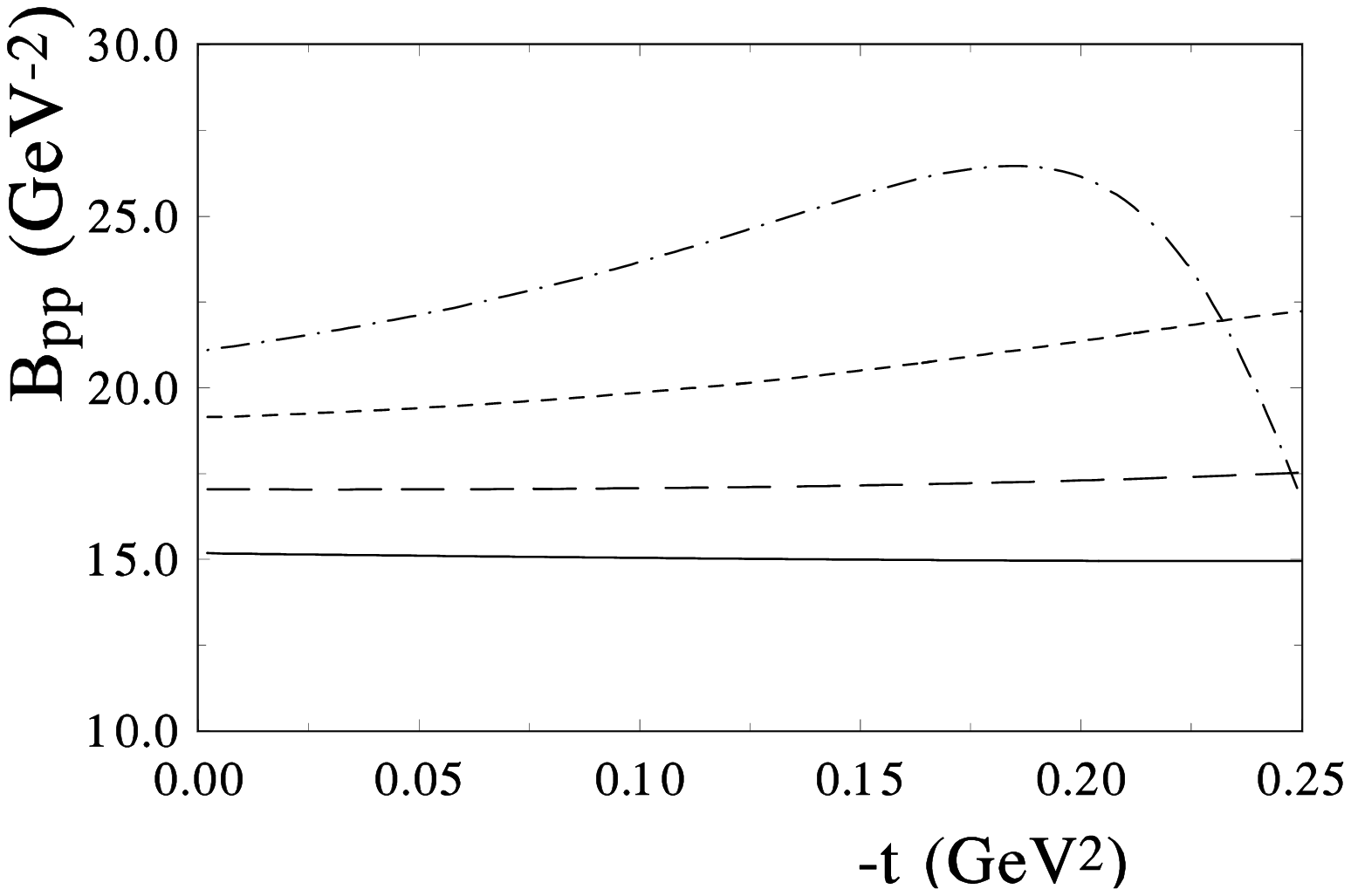}
\caption{Results of the ETPM model: 
$\rho(s,t)$ (upper panel) and $B(s,t)$ (lower panel)
at $100$ GeV (plain curve), $500$ GeV (long dashes), 
5 TeV (short dashes) and 14 TeV (dash-dotted curve).}
\end{figure}
We can now explore the consequences of these effects on the experimental
measurement itself. We shall compare in the following the situation at
2 TeV with that at 14 TeV. We insist that the curves we give for the cross sections
and for the $\rho$ parameter are only illustrative of an effect present in many models.
The essential ingredients are a sizeable value of $\rho$, and the strong dependence of 
$\rho$ and $B$ on $t$ once the black-disk limit is reached.

\section{Fitting procedure with luminosity-dependent method  }
The number of elastic events is related to the total hadronic cross 
section through the following formula:
\begin{eqnarray}
\frac{dN}{dt}&&= {\cal{L}} \left[\frac{4 \pi \alpha^2}{|t|^2} G^4(t)\right. 
\nonumber \\
&  & - \frac{2 \alpha\left(\rho(s,t) + \phi_{CN}(s,t)\right) \sigma_{tot}G^2(t) e^{-\frac{B(s,t)|t|i}
{2}}}{|t| }       \nonumber \\
     &  &\left. +\frac{\sigma_{tot}^2(1+\rho(s,t)^2)e^{-B(s,t)|t|}}{16 \pi}\right] 
\label{dN/dt}
 \end{eqnarray}
 where  ${\cal{L}}$ is the luminosity, the first term is the Coulomb term ($\alpha$ is the electromagnetic
coupling constant and $G(t)$ the electromagnetic form factor given by
$
G^2(t) = \left(4 m_p^2 - \mu t \right)\left(4 m_p^2-t\right)/\left(\Lambda^2 (\Lambda - t)^2\right)
$ with $m_p$ the proton mass, $\Lambda=0.71$ GeV$^2$ and $\mu=2.79$), the second term is the interference term between the Coulomb amplitude and the hadronic amplitude ($\phi_{CN}(s,t)$ is the phase of the Coulomb-Nucleon Interference (CNI) term \cite{SelyuginMP96}), from which one can extract $\rho$, and the third term is the purely hadronic contribution.

\begin{table}[htp]
\label{tab:4}       
\begin{center}
\begin{tabular}{|c|c|c|c|c|}
 \multicolumn{5}{c} {input}   \\ 
\hline 
$\sqrt{s}$ &$\cal L$ (fb$^-1$) &  $\sigma_{tot} $ (mb)   & $ \rho(s,0)$    &    B(s,0) (GeV$^{-2}$) \\ \hline
 2 TeV&  1 &  $82.7 $  &  $  0.23 $  &18.7  \\
 14 TeV&1   &  $152.5 $    &  $  0.24 $   &  21.4    \\
\hline
\multicolumn{5}{c} {output for $\cal L$ and $\rho$ fixed }   \\ \hline
  2 TeV& $ 1  $ & $ 83.61 \pm .44 $  &  $  0.15$
  & $ 23.6 \pm 0.2 $  \\
  2 TeV& $ 0.95 $ & $ 85.8 \pm .45 $  &  $  0.15 $
  & $ 23.6 \pm 0.2 $  \\
\hline
\multicolumn{5}{c} {output for all parameters free }   \\
\hline
2 TeV&  $ 0.93 \pm 0.07 $ & $ 85.2 \pm 3 $  &  $  0.15 \pm 0.04 $
  & $ 18.1 \pm 0.25 $  \\

14 TeV&  $ 1.15 \pm 0.05 $ & $ 142.3 \pm 2.8 $  &  $  0.29 \pm 0.06 $
  & $ 23.6 \pm 0.2 $  \\
 \hline
\end{tabular}
\caption{Input parameters for the simulated data  at $\sqrt{s} = 2 \ $TeV and  $\sqrt{s} = 14 \ $TeV obtained in EPTM model, and the results of fits to these data
 with a simple exponential form of the scattering amplitude.
}
\end{center}
\end{table}
\begin{figure}[htp]
\includegraphics[width=0.4\textwidth] {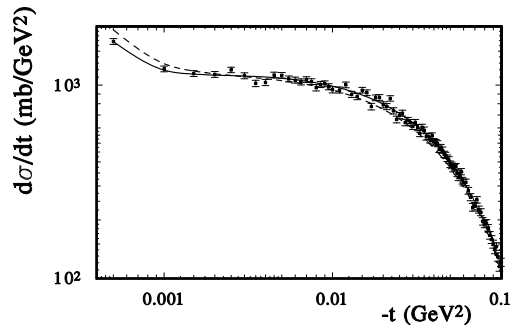}
\caption{The simulated data
at $\sqrt{s}= 14 \ $TeV, the theoretical curve from which the data are generated (plain) and the  fit to them for $\rho$ fixed at 0.1 (dashed).
}
\end{figure}
\begin{figure}[htp]
\includegraphics[width=0.4\textwidth] {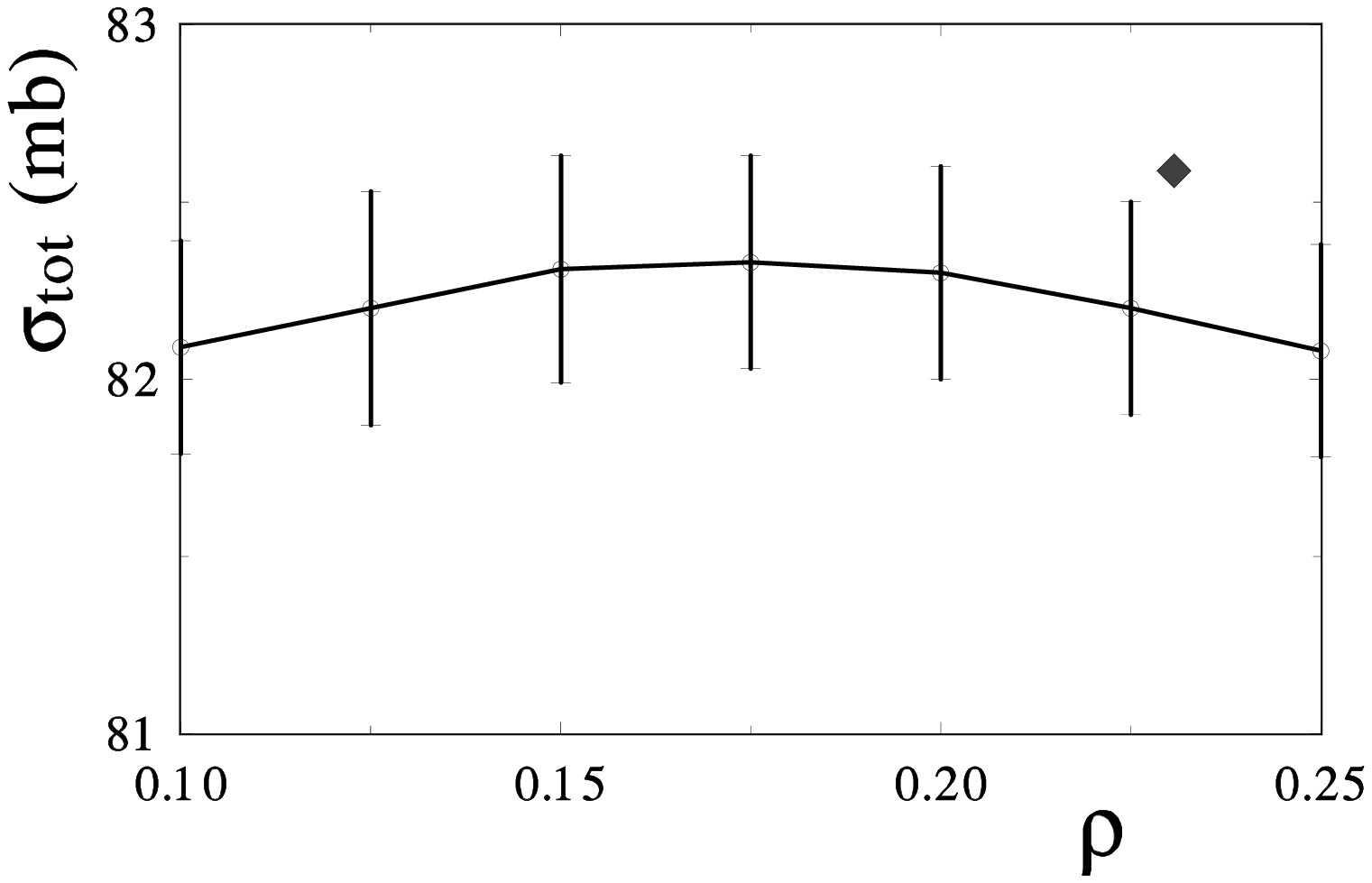}
\vglue -1.cm
\includegraphics[width=0.4\textwidth] {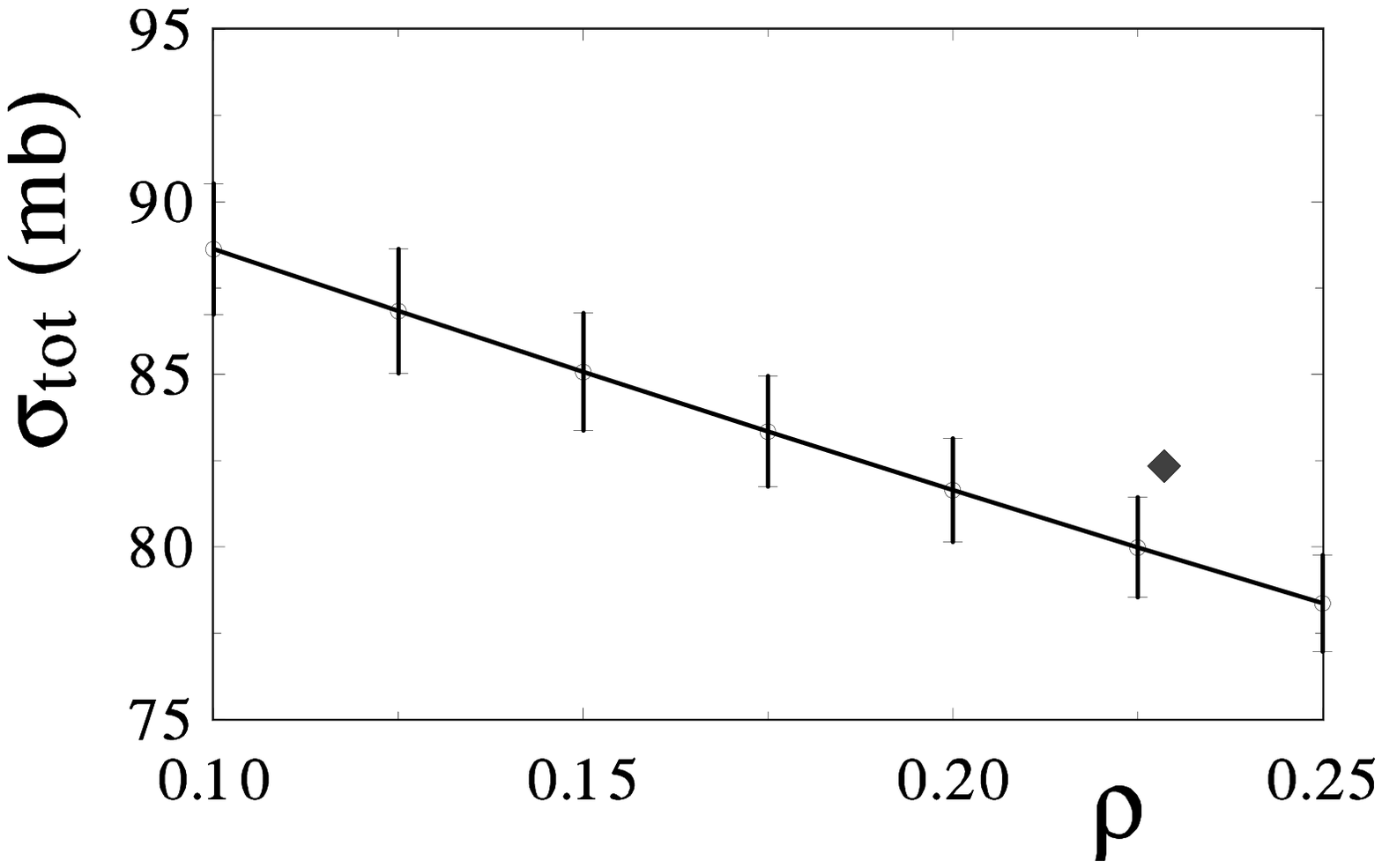}
\caption{ The size of $\sigma_{tot}$ obtained by fitting the simulated data assuming 
a fixed value of $\rho$ at $\sqrt{s}=2 \ $TeV    with ${\cal L}$ fixed at the input value (upper panel) and with 
free $\cal L$ (lower panel); the diamond gives the true value of  $\sigma_{tot}$ 
and $\rho(s,0)$.}
\end{figure}

We can use this formula to generate simulated data  for two energies
 $\sqrt{s} =2 \ $TeV and   $\sqrt{s} =14 \ $TeV, using $B(s,t)$ and $\rho(s,t)$ calculated in the ETPM. 
We assume that  $90$ points will be measured in a $t$ 
interval identical to that of the UA4/2 experiment $ -0.1 {\rm\ GeV}^2\leq t\leq -0.0006 {\rm\ GeV}^2 $ \cite{UA42}.
We then randomize the theoretical (``true") curve assuming Gaussian errors similar to those 
of UA4/2. The resulting simulated data are shown in Fig. 2 for $\sqrt{s}=14$ TeV,
and correspond to the parameters given in Table 1. One can 
then fit these simulated data according 
to Eq.~\ref{dN/dt} but assuming constant $B$ and $\rho$. One has 2 extra 
parameters besides $\rho$ and $B$: $\cal L$, the luminosity, and the total cross section 
$\sigma_{tot}$, which is what one aims to measure. 

The result of the fitting procedure at 2 TeV is shown in Fig. 3, 
where the correlation between 
the value of $\rho$ and $\sigma_{tot}$ is shown.
If $\cal L$ is  fixed at $1$ fb$^{-1}$ ({\it i.e.} if we know the luminosity),
 the difference between the central values of the fitted
$\sigma_{tot}$ is small $\approx 0.3 \ $mb; the errors from the fitting procedure 
are $1 \ $mb and the obtained  value of $\sigma_{tot}$
differs from the input by $0.5 \ $mb.
A very different picture appears  if we  fit the luminosity. 
We obtain then an error $\Delta\sigma_{tot} = 1.9 \ $mb
and the correlation between  $\sigma_{tot}$ and the assumed value of $\rho$
is very high, as seen in Fig. 3 and Table 1.  The result of a joint fit to $\cal L$, $B$, $\sigma_{tot}$ and $\rho$ is also shown in Table 1. All parameters
are within 1 $\sigma$, except $\rho(0)$ which is $2 \sigma$ from the input.

The situation at $\sqrt{s} = 14 \ $TeV is much worse:    
one sees from Fig. 4 that the result of the  correlation between
 $\rho$ and $\sigma_{tot}$ increases drastically: even if one knows the luminosity,  
 the dependence of $\sigma_{tot}$ on $\rho$ is very strong. 
The difference between the true $\sigma_{tot}$  and the fitted value reaches 
$2.5 \ $mb, as shown in Table 1. All the parameters
are now several standard deviations from their true value. Note also that, because the CNI term is negative, a decrease in $\rho$ from its true value 0.24 to 0.1 leads to an increase in the value of $d\sigma/dt$, as seen in Fig. 2.

\section{Fitting procedure with the luminosity-independent method  }
Another way to extract the total cross section 
(for example, see \cite{Abe1}) 
 is the luminosity-independent method, which gives
\begin{equation}
\sigma_{tot}  =
\frac{16 \pi}{1+\rho^2}\frac{(dN_{el}/dt)|_{t=o}}{N_{el} +N_{inel} },
\label{lum-ind}
\end{equation}
To obtain the information needed for this method
one must measure the elastic rate at values of $|t|$ large enough to
neglect the Coulomb amplitude \footnote{otherwise one must subtract the 
electromagnetic 
and CNI terms, but this means that one should know
the luminosity beforehand}.    
Hence, one hopes to obtain a more accurate result with less information.

This method relies on the hope that $N_{el} + N_{inel}$ can be measured 
accurately. However, there are three problems. The first one concerns the Coulomb and CNI
regions:  one needs to cut them off, but $N_{el} + N_{inel}$ is
the total number of hadronic events, so one must
compensate somehow for that cut. The second problem comes from the fact that 
part of the inelastic events (such as $N^*$ production) will escape the detector.
The third problem, which we shall address here, concerns the extrapolation of 
 $dN/dt$ from a minimum $|t|$  far away from the CNI region to $t=0$.
 
To simulate this analysis, we take $N_{el}+N_{in}=n\sigma_{tot}$, and let $n$
go from 0.9 to 1.1.
We adopt the cuts planned for the TOTEM experiment at the LHC,
 $-t$ in $[0.03, 0.1]$ GeV$^2$ , so we are left with in 51
 simulated data points.

For the analysis at 2 TeV, if we take $n=1$ the dependence of $\sigma_{tot}$ over 
$\rho$ will be comparable to that shown in  Fig.~2, but the errors on $\sigma_{tot}$ 
will increase. For example, if we fix $\rho=0.15$, then $\sigma_{tot}= 83.61 \pm 0.44 \ $mb. $\sigma_{tot}$ changes by $1.9 \ $mb when $\rho$ goes from $0.05$ to $0.25$. If $n$ can go 
as low as 0.95, then the measurement of $\sigma_{tot}$ increases, {\it e.g.} 
$\sigma_{tot}=85.8$ mb when $\rho=0.15$. 

At $\sqrt{s} =14 \ $TeV, the changes are more pronounced, as seen in Fig. 4.
Cutting off the CNI region removes the possibility to measure $\rho$. As the
normalization is inversely proportional to $1+\rho^2$, it is rather obvious that
if one allows $\rho$ to range from 0.05 \cite{COMPETE} to 0.3, one will get
a 10 \%
change in $\sigma_{tot}$, which will only be added to the uncertainty coming from
the estimate of $N_{el} + N_{inel}$. So it seems to us that it is illusory to hope for 
an accuracy on $\sigma_{tot}$ better than 10 \% from this method.
\begin{figure}[htp]
\includegraphics[width=0.4\textwidth] {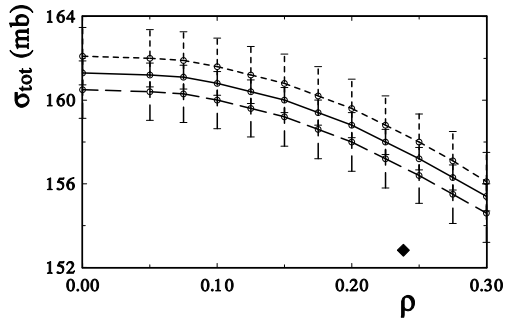}
\caption{$\sigma_{tot}$  
at $\sqrt{s}=14 \ $TeV  with fixed normalization
($n=1$) (central curve) and $n= 1\pm 0.1$ (exterior curves). The diamond 
indicates the input $\sigma_{tot}$ and $\rho(s,0)$ from which the data were simulated.
}
\end{figure}

The large discrepancy between the measurements of $\sigma_{tot}$ by CDF \cite{Abe1} and E710/E811 \cite{E710} probably has its origin in the uncertainty on $N_{el}+N_{in}$ and on the dependence of $\rho$ and $B$ on $t$, and not in some experimental mistake.
Hence this large difference really reflects the real error one is to expect from this method. 
\section{Conclusion} 
Our analysis shows that both methods introduce large correlations between $\rho$ 
and $\sigma_{tot}$. As it is very likely that at the LHC unitarisation will play an 
important role, one should not assume that  $B(s,t) $ and  $ \rho(s,t) $ are constant 
with $t$, and their exact behavior with $t$ is model-dependent. Hence the 
inescapable conclusion is that a 1~\% measurement of $\sigma_{tot}$ will only be 
possible if one measure $B(s,t)$ and $\rho(s,t)$ as well.

 $|t|$ should go from very small values, as close as possible to zero, to
about 0.1 $GeV^2$, with sufficiently small bins. And it will be important to allow all 
parameters 
to vary. Indeed, the measurement of $\rho$ performed by UA4/2 ($0.135 \pm 
0.015$) \cite{UA42} seemed to contradict that of UA4 ($0.24 \pm 0.02$) \cite{ua4} 
only because $\sigma_{tot}$ was fixed. Allowing $\sigma_{tot}$ to be fitted to leads 
to an agreement between the two measurements \cite{PL94}.

We also believe that the luminosity-dependent method is preferable, as it uses more 
information. The measurement of $\rho$ performed at the Tevatron \cite{Amos-rho} 
used the luminosity-independent method with very large bins in $t$, and the 
interval considered was $0.00095  \leq |t| \leq 0.1431 \ $GeV$^2$. On the lower 
side, one reached very small $t$, so that the behavior of the amplitude cannot be 
taken as a single exponential because of the CNI effect. On the upper side, the 
intervals in $t$ were too big to measure  the  specific properties of the CNI region. 
Hence the $\rho$ parameter extracted is very uncertain, and it could be that it varied 
appreciably with $t$.  

As the standard fitting procedure can give misleading results, we remind the reader 
that additional methods have been proposed to check the validity of the assumptions 
entering the fits. Firstly, it is possible to extract the value of $\rho(s,t)$ at small but nonzero $t$ \cite{SelyuginBL95,SelyuginPL05}, using the fact that the Coulomb amplitude $F_C(t)$ has an opposite sign to that of the real part of the $pp$ 
amplitude $F_n(t)$, so that there is a value $t=t_C$ for which $F_C(t_C)=-\Re e F_n(t_C)$, so that the differential cross section $d\sigma/dt$ has a local minimum at that value. The position of this minimum depends strongly on the form assumed for $\rho(s,t)$ and extracting its value would show
whether $\rho$ varies quickly with $t$ or not. Using this method, it was found in 
\cite{SelyuginBL95,SelyuginPL05} that already at $\sqrt{s}=52.8$~GeV, $\rho$ was not constant with $t$.

It is also possible, at small $|t|$, to determine the elastic hadronic cross section via an iterative method, which takes advantage of the expression of the total elastic cross section in the CNI region \cite{SelYad}.
 
Finally, one can adapt a method that was first designed to study eventual 
oscillations in $d\sigma/dt$ \cite{SelyuginOSC}. 
The idea is to compare two statistically
independent samples built by binning the whole $t$ range in small intervals,
 and by keeping {\it e.g.} one interval out of two.
The deviations of the
experimental values from theoretical expectations, weighted by the experimental 
error,
are then summed for each sample $k$:
\begin{equation}
\Delta R^k(t)= \sum_{|t_i|<|t|}[(d\sigma^k/dt_{i})^{exp}
  -  (d\sigma/dt_{i})^{th}] / \delta_{i}^{exp},
\end{equation}
 where $\delta_{i}^{exp} $ is the experimental error.
 If the theoretical
curve does not precisely describe the experimental data,
(for example, if the physical
hadron amplitude does not have an exactly
exponential behavior with momentum transfer), the sum $\Delta R^k(t) $
will differ from zero, going beyond the size
of the statistical error.

Using these methods will help to test the assumptions entering the future
experimental analyses of TOTEM and ATLAS, and may lead to a much more reliable
measurement of $\rho$ and $\sigma_{tot}$.
\acknowledgments{
 O.V.S. gratefully acknowledges financial support
  from FRNS and would like to thank the  University of Li\`{e}ge
  where part of this work was done.
    }

\end{document}